\documentclass[aps,prb,twocolumn,showpacs,superscriptaddress,bibnotes]{revtex4}

\usepackage{graphicx}
\usepackage{dcolumn}
\usepackage{bm}

\begin{document}

\title{Prominent 5d-orbital contribution to the conduction electrons in gold}

\author{A. Sekiyama}
\affiliation{Division of Materials Physics, Graduate School of Engineering Science, 
Osaka University, Toyonaka, Osaka 560-8531, Japan}
\affiliation{SPring-8/Riken 1-1-1Kouto, Mikazuki, Sayo, Hyogo 679-5148, Japan}
\author{J. Yamaguchi}
\affiliation{Division of Materials Physics, Graduate School of Engineering Science, 
Osaka University, Toyonaka, Osaka 560-8531, Japan}
\author{Higashiya}
\affiliation{SPring-8/Riken 1-1-1Kouto, Mikazuki, Sayo, Hyogo 679-5148, Japan}
\affiliation{Industrial Technology Center of Wakayama Prefecture, 
Wakayama 649-6261, Japan}
\author{M. Obara}
\author{H. Sugiyama}
\author{M. Y. Kimura}
\affiliation{Division of Materials Physics, Graduate School of Engineering Science, 
Osaka University, Toyonaka, Osaka 560-8531, Japan}
\author{S. Suga}
\affiliation{Division of Materials Physics, Graduate School of Engineering Science, 
Osaka University, Toyonaka, Osaka 560-8531, Japan}
\affiliation{SPring-8/Riken 1-1-1Kouto, Mikazuki, Sayo, Hyogo 679-5148, Japan}
\author{S. Imada}
\affiliation{Department of Physical Sciences, Ritsumeikan University, 
Kusatsu, Shiga 525-8577, Japan}
\author{I. A. Nekrasov}
\affiliation{Institute of Electrophysics, Russian Academy of 
Sciences-Ural Division, 626041 Yekaterinburg, GSP-170, Russia}
\author{M. Yabashi}
\affiliation{SPring-8/Riken 1-1-1Kouto, Mikazuki, Sayo, Hyogo 679-5148, Japan}
\affiliation{SPring-8/JASRI 1-1-1 Kouto, Mikazuki, Sayo, Hyogo 679-5198, Japan}
\author{K. Tamasaku}
\author{T. Ishikawa}
\affiliation{SPring-8/Riken 1-1-1Kouto, Mikazuki, Sayo, Hyogo 679-5148, Japan}

\begin{abstract}
We have examined the valence-band electronic structures of gold and silver 
in the same column in the periodic table with nominally filled d orbitals 
by means of a recently developed polarization-dependent hard x-ray 
photoemission. 
Contrary to a common expectation, it is found that the 5d-orbital 
electrons contribute prominently to the conduction electrons in gold 
while the conduction electrons in silver are to some extent free-electron-like 
with negligible 4d contribution, 
which could be related to a well-known fact that gold is more stable than silver in air. 
The 4d electron correlation effects are found to be essential for the conduction electron 
character in silver. 

\end{abstract}

\date{\today}


\maketitle

\section{Introduction}
Single-element materials in the same column in the periodic table often show 
mutually similar features in such cases of alkaline (-earth) metals 
and halogens. 
On the other hand, it is known for noble metals with group 11 
in the periodic table that gold has 
considerably different chemical stability from that of silver, 
although their Fermi surface topology~\cite{dHvAAg,dHvAAu} 
and predicted band dispersion near the Fermi level ($E_F$) are 
mutually similar~\cite{Auband2,Auband1,Agband1,Jepsen1}. 
For instance, it is well known that solid gold is very stable 
under many circumstances while solid silver is gradually oxidized in air. 
The valence-band electronic configurations of these solids per atom 
have so far been recognized~\cite{Auband1,Agband1,PES1970} 
to be composed of fully occupied $n$d states ($n$d$^{10}$) 
plus one conduction electron occupying an $(n+1)$sp state 
($n = 4$ for silver and $n = 5$ for gold) 
although these configurations have not been experimentally verified to date. 
In this paper, however, we show that the 5d-orbital contribution 
to the conduction electrons is actually prominent in bulk gold 
while the above configuration is practically correct for bulk silver, 
by means of the linear polarization-dependent hard x-ray photoemission 
which we have very recently developed. 
These findings could be related to a well-known fact that 
gold is more stable than silver in air. 

There are few experimental techniques to probe the orbital contributions 
in the valence-band electronic states of solids. 
Such spectroscopic technique as resonance 
photoemission~\cite{AllenAP1986,Fujimorid1PES,N2000} 
and surface-sensitive low-energy two-dimensional angle-resolved 
photoemission over wide emission angles~\cite{D2PES} 
are not practical at all 
for bulk polycrystalline gold and silver to quantitatively 
clarify the itinerant s, p and d orbital contributions 
to the conduction electrons. 
Rough information of the orbital contributions can be obtained 
from a comparison of angle-integrated valence-band photoemission spectra 
at considerably different excitation 
energies~\cite{HAXPESAg,MatsunamiYb08} 
such as $h\nu \sim$1 and $\sim$8 keV, 
for which the relative photoionization cross-sections 
depend on $h\nu$~\cite{Scofield,Lindau}. 
However, the $h\nu$ dependence of the spectra is often seen 
also by the different bulk/surface contribution ratio 
due to the different photoelectron kinetic 
energies~\cite{ASPRL2004}, 
which prevents us to reliably estimate the orbital contributions. 

At hard x-ray excitations, the photoelectron cross-sections 
for the s and p states per an electron become comparable to 
those for the d and f states 
whereas they are very small at soft x-ray 
excitations~\cite{Scofield,Pdep1,Pdep2,Pdep3}. 
This is also the case for gold and silver. 
Moreover, it has been predicted theoretically that 
the photoelectron angular distribution with respect to the angle $\theta$ 
between the two directions of photoelectron detection 
and the linear polarization (electric field) of incident light has 
strong orbital dependence. 
As an overall tendency, the calculation~\cite{Pdep1,Pdep2,Pdep3} 
predicts that 
a ratio of the photoelectron intensity toward the direction 
perpendicular to the polarization vector 
($\theta = 90^{\circ}$) to that along the polarization vector 
($\theta = 0^{\circ}$), 
defined as $I_{\theta = 90^{\circ}}/I_{\theta = 0^{\circ}}$, 
is very small as $\leq \sim$0.1 for the s and $i$p ($i > 4$) 
states compared with that ($\geq$ 0.2) for the d and f states 
at $h\nu$ = 5-10 keV. 
When we measure the hard x-ray valence-band 
photoemission~\cite{HXPES2003,YamaSmOsSb,YbB122009} spectra 
with different linear polarization, 
a photoelectron intensity ratio 
$I_{\theta = 90^{\circ}}/I_{\theta = \theta '}$ (hereafter simply called ``ratio") 
practically equivalent to $I_{\theta = 90^{\circ}}/I_{\theta = 0^{\circ}}$ 
($\theta '$ should be much deviated from 90$^{\circ}$ as  
in our present experimental geometry with 30$^{\circ}$, 
see the inset of Figure~\ref{AuCore}) can be obtained. 
Therefore, the extraction of the s and $i$p contributions 
as well as that of the d and f contributions 
in the bulk valence band, 
for which there has been no conclusive experimental technique to date 
as mentioned above, become feasible 
by the linear polarization-dependent hard x-ray photoemission measurements. 

\section{Experimental}

Polarization-dependent hard x-ray photoemission measurement was 
performed at BL19LXU of 
SPring-8~\cite{YabashiPRL01} with a 27-m long undulator 
by using a newly developed MBS A1-HE hemispherical photoelectron spectrometer. 
A Si 111 double-crystal monochromator selected $\sim$8 keV radiation 
with linear polarization along the horizontal direction 
(the so-called degree of linear polarization $P_L > +0.98$), 
which was further monochromatized by a channel-cut crystal with the Si 444 reflection. 
In order to switch the linear polarization of the hard x-ray from the horizontal 
to vertical directions, a (100) diamond was used as a phase retarder~\cite{MotoDia} 
in the Laue geometry with the 220 reflection, 
which was placed downstream of the channel-cut crystal. 
The transmittance of the x-ray at $\sim$8 keV for the diamond was confirmed 
as $\sim$35\%. 
$P_L$ of the x-ray downstream from the phase retarder was estimated as 
about $-0.8$, which corresponds to the linear polarization components 
along the horizontal and vertical directions of 10\% and 90\%, respectively. 
As shown in the inset of Figure~\ref{AuCore}, 
the emission direction of photoelectron 
to be detected was set within the horizontal plane, 
where $\theta$ between the light polarization and the electron emission angle 
was 30$^{\circ}$ (90$^{\circ}$) at the excitation with the horizontally (vertically) 
linear polarization. 

The polycrystalline gold and sliver prepared by {\it in situ}  evaporation 
were measured at 12-15 K, 
where the base pressure was $\sim$8 x 10$^{-8}$ Pa. 
The energy resolution was set to $\sim$280 ($\sim$400) meV 
for the measurement of silver (gold). 
The absence of the O 1s and C 1s spectral weight was confirmed. 
The spectral weights were normalized by the photon flux. 
It should be noted that this technique is useful even for polycrystalline samples 
as we demonstrate in this paper  
since the angular distribution depends mainly on $\theta$, 
which is determined by the measurement geometry.

\section{Polarization dependence of the core-level spectra for gold}

\begin{figure}
\includegraphics[width=8.5cm,clip]{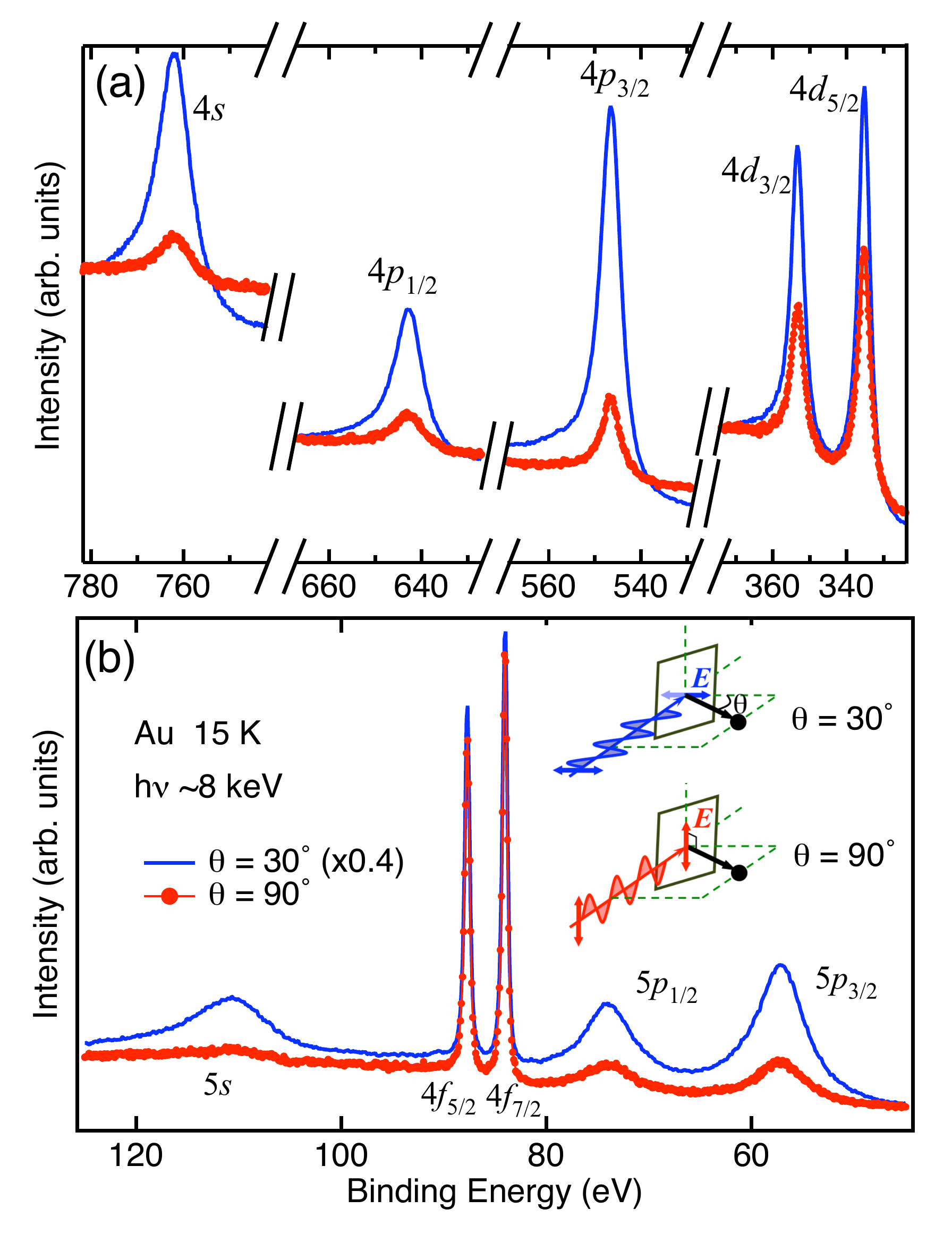}
\caption{
(a) Linear polarization dependence of the hard x-ray excited 4s, 4p and 4d 
core-level photoemission spectra of polycrystalline gold. 
(b) Same as (a) but of the shallow 5s, 4f and 5p core-level spectra. 
The experimental geometries and their notations with respect to the directions 
of light propagation, polarization (electric field) vector {\bf {\it E}} 
and detected photoelectrons are shown in the inset, 
in which the black circles denote the photoelectrons. 
The spectral intensity at $\theta$ = 30$^{\circ}$ is scaled by multiplying a factor of 0.4 
for comparison with that at $\theta$ = 90$^{\circ}$. 
}
\label{AuCore}
\end{figure}

We show the polarization dependence of the 
core-level photoemission spectra for polycrystalline gold in Figure~\ref{AuCore}. 
The photoemission spectral weight at $\theta$ = 90$^{\circ}$ is 
more strongly suppressed compared with that at $\theta$ = 30$^{\circ}$ 
for the 4s, 4p, 5s and 5p core levels than for the 4f levels. 
The intensity ratios $I_{\theta = 90^{\circ}}/I_{\theta = 30^{\circ}}$ 
for the core levels estimated from our experimental data are consistent 
with the calculation~\cite{Pdep1,Pdep2,Pdep3} as shown in Table~\ref{ratio} 
except for the s states. 
The experimentally estimated ratios for the 4s and 5s core levels 
are much larger than the predicted values from the calculation, 
but still smaller than those of the 4d and 4f states. 

\begin{table*}
\caption{\label{ratio}Intensity ratio 
$I_{\theta = 90^{\circ}}/I_{\theta = 30^{\circ}}$ 
for the gold core-level excitations at the kinetic energy of $\sim$8 keV. 
The parameters for the calculations are listed in References~\onlinecite{Pdep2,Pdep3}. 
For the calculations, see also Appendix}
\begin{ruledtabular}
\begin{tabular}{ccccccccccc}
 &4s&4p$_{1/2}$&4p$_{3/2}$&4d$_{3/2}$&4d$_{5/2}$&4f$_{5/2}$&4f$_{7/2}$&5s&5p$_{1/2}$&5p$_{3/2}$\\
\hline
Experiment&0.07&0.11&0.09&0.2&0.24&0.4&0.4&0.1&0.16&0.13\\
Calculation&0.02&0.1&0.07&0.24&0.3&0.54&0.54&0.02&0.1&0.07\\
\end{tabular}
\end{ruledtabular}
\end{table*}

\section{Polarization-dependent valence-band spectra of silver and gold}

\begin{figure}
\includegraphics[width=8cm,clip]{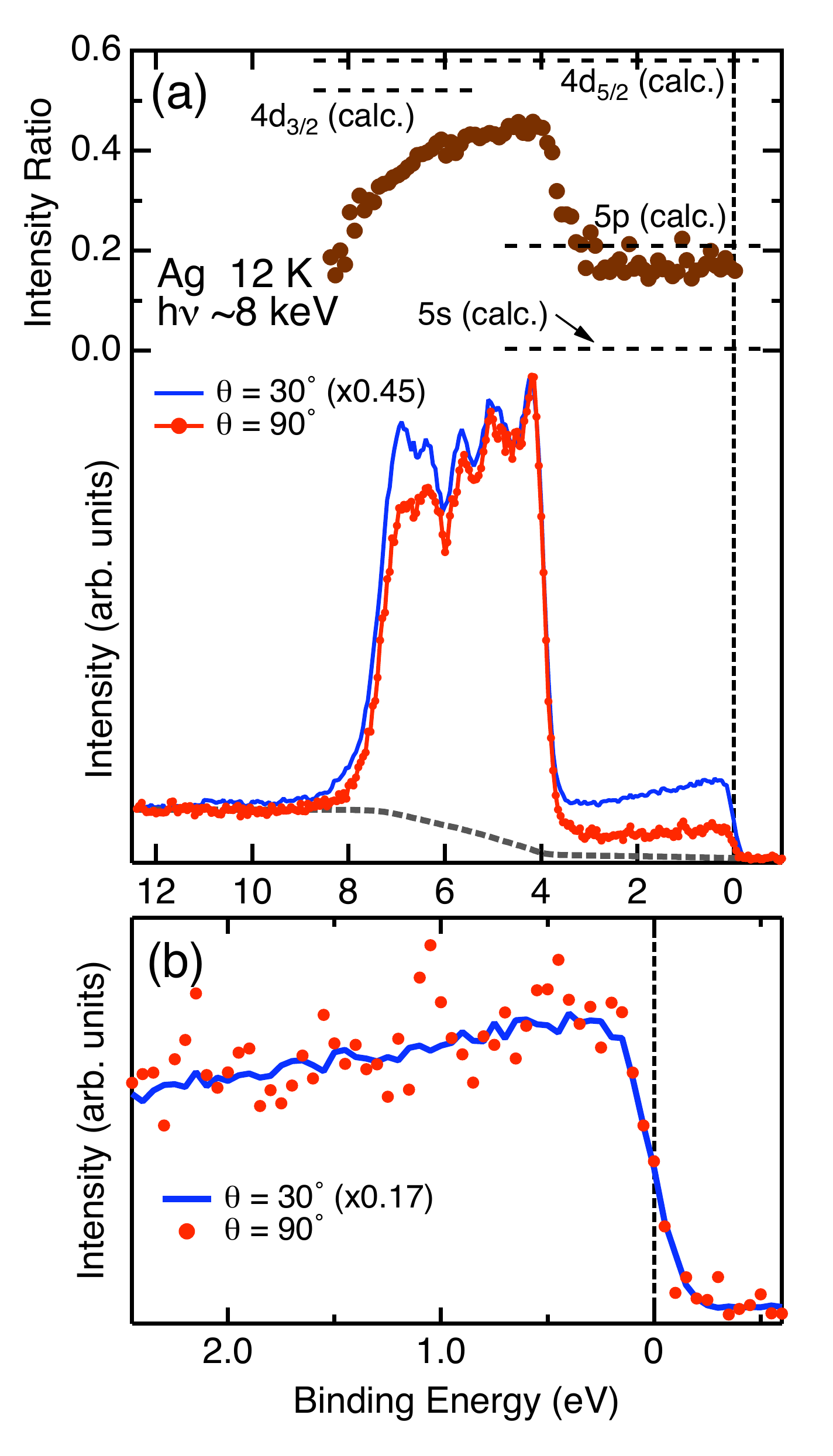}
\caption{
Linear polarization dependence of the valence-band spectra of silver.  
(a) Spectra in a whole valence-band region. 
The spectral intensity at $\theta$ = 30$^{\circ}$ is scaled by multiplying 0.45 
for direct comparison with that at $\theta$ = 90$^{\circ}$. 
A Shirley-type background is shown by a grey dashed-line. 
The top graph represents the ratio $I_{\theta = 90^{\circ}}/I_{\theta = 30^{\circ}}$. 
The expected ratios for the 4d$_{3/2}$, 4d$_{5/2}$, 5s and 5p states 
from the calculation~\cite{Pdep1,Pdep3} are shown by dashed horizontal lines.  
For the 5p state, we have used the calculated ratio for the In 5p state 
since there is no calculation for the "Ag 5p level". 
The ratio $I_{\theta = 90^{\circ}}/I_{\theta = 30^{\circ}}$ was obtained by the spectral weight 
at $\theta$ = 90$^{\circ}$ divided by that at $\theta$ = 30$^{\circ}$ after subtracting the 
Shirley-type background.
 It should be noted that this ratio hardly changes between $E_F$ and 5 eV 
 irrespective of the background-subtraction procedure. 
(b) Spectra near $E_F$, where the spectral intensity at $\theta$ = 30$^{\circ}$ 
is scaled by multiplying 0.17 
for direct comparison with that at $\theta$ = 90$^{\circ}$.}
\label{AgVal}
\end{figure}

Figure~\ref{AgVal} shows the polarization dependence of the valence-band spectra 
for polycrystalline silver. 
There is a strong spectral weight between 4 and 7 eV in both spectra 
at $\theta$ = 30$^{\circ}$ and 90$^{\circ}$, 
which is predominantly ascribed to the 4d contributions. 
Whereas this feature is consistent with the previous photoemission 
studies~\cite{PES1970,HAXPESAg}, 
it is found that the experimentally estimated 
$I_{\theta = 90^{\circ}}/I_{\theta = 30^{\circ}}$ in this energy region 
is consistent with the calculated ratio for the 4d excitations 
as shown in the upper panel of Figure~\ref{AgVal}(a). 
The experimental ratio decreases rapidly from $\sim$4 to $\sim$3 eV 
and then stays almost flat toward $E_F$, indicating that 
the spectral weight near $E_F$ is strongly suppressed 
at $\theta$ = 90$^{\circ}$ compared with that of the 4d states 
in the region of $4-7$ eV. 
The intensity ratio near $E_F$ is larger than the calculated ratio 
for the 5s state and slightly less than that for the 5p state. 
The slope of the intensity from 2.5 eV to $E_F$ 
at both $\theta$ = 30$^{\circ}$ and 90$^{\circ}$ is qualitatively 
consistent with that of the partial density of states (PDOS) 
with s and p symmetries, 
which has been obtained by our band-structure (local density 
approximation, LDA) calculation 
by using the WIEN2k package~\cite{Wien2k}, 
but incompatible with that of the PDOS with d symmetry. 
These results reveal that the 4d bands are located far below $E_F$ 
and well separated from the conduction 5sp band in an energy region 
from $E_F$ to $\sim$3 eV. 
It is thus experimentally confirmed that the 4d orbitals are nearly fully occupied 
in the solid silver as expected for a long time, 
which has not been unfortunately verified by the previous high-energy 
photoemission~\cite{HAXPESAg}. 
The reduction of $I_{\theta = 90^{\circ}}/I_{\theta = 30^{\circ}}$ around 6.5 eV 
compared to that at 4.5 eV is mainly due to the mixture of the 5s state, 
which has been suggested by the previous band-structure 
calculations~\cite{Agband1,Jepsen1} 
as well as our band-structure calculation.

\begin{figure}
\begin{center}
\includegraphics[width=8cm,clip]{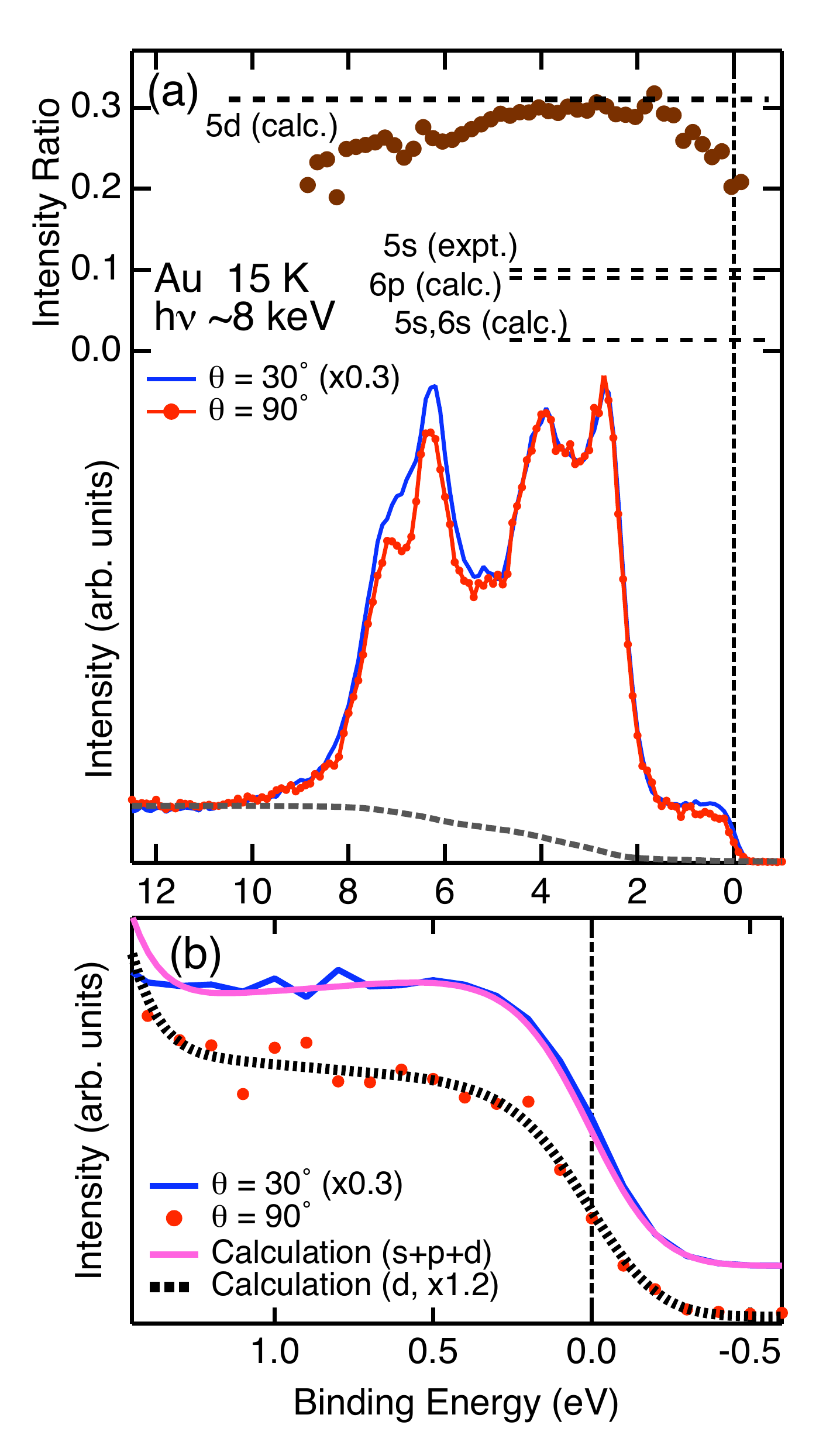}
\end{center}
\caption{
Linear polarization dependence of the valence-band spectra of gold.  
(a) Same as Figure~\ref{AgVal}(a) but for polycrystalline gold. 
The spectral intensity at $\theta$ = 30$^{\circ}$ is scaled by multiplying 0.3 
for comparison with that at $\theta$ = 90$^{\circ}$. 
The expected intensity ratios for the 5d, 6s and 6p states 
from the calculation~\cite{Pdep2,Pdep3} are shown by horizontal dashed lines.  
Since there is no calculation for the Au 6p level, 
we have used the calculated ratio for the Tl 6p state. 
(b) Spectra near $E_F$ at $\theta$ = 30$^{\circ}$ and 90$^{\circ}$ in comparison 
with the sum of s, p and d partial density of states, 
and the only d partial density of states obtained by our band-structure calculation, 
respectively.
The partial densities of states and its sum were broadened 
by the experimental resolution after multiplying the Fermi-Dirac function 
at the measuring temperature of 15 K. 
}
\label{AuVal}
\end{figure}

We next show that the polarization dependence of the valence-band spectra 
for bulk gold is not only quantitatively but also qualitatively different 
from that for silver, as demonstrated in Figure~\ref{AuVal}. 
The intensity ratio $I_{\theta = 90^{\circ}}/I_{\theta = 30^{\circ}}$, 
which hardly changes from 5 to 2 eV in the 5d band region, 
decreases gradually from 2 eV toward $E_F$ 
without showing a rapid suppression. 
The ratio $I_{\theta = 90^{\circ}}/I_{\theta = 30^{\circ}}$ 
in the vicinity of $E_F$ estimated as $\sim$0.22 is much larger than 
the calculated values for the 6s and 6p states. 
Even if $I_{\theta = 90^{\circ}}/I_{\theta = 30^{\circ}}$ for the 6s state is 
in fact larger than the calculated value and close to 
the experimentally obtained ratio as $\sim$0.1 for the 4s or 5s 
core-level state, 
$I_{\theta = 90^{\circ}}/I_{\theta = 30^{\circ}}$ in the vicinity of $E_F$ is 
still much larger than these values. 
In addition, the spectral line shape from $E_F$ to 1.5 eV is different 
between $\theta$ = 30$^{\circ}$ and 90$^{\circ}$ 
as shown in Figure~\ref{AuVal}(b), 
which is in contrast to that for silver in Figure~\ref{AgVal}(b). 
While the intensity is nearly flat or slightly enhanced from 1.5 eV to $E_F$ 
at $\theta$ = 30$^{\circ}$, 
it is gradually reduced toward $E_F$ at $\theta$ = 90$^{\circ}$. 
The different slope of the intensity for different $\theta$ suggests that 
the 5d-orbital contribution is extended into the region of $E_F - 2$ eV 
and the additive 6sp-orbital components contribute in the spectrum 
at $\theta$ = 30$^{\circ}$. 

The polarization-dependent spectra of gold near $E_F$ are consistent 
with the result of the band-structure calculation. 
As shown in Figure~\ref{AuVal}, the spectrum at $\theta$ = 90$^{\circ}$ is 
excellently reproduced by the d PDOS 
whereas a naive sum of the s, p and d PDOS well simulates the spectrum 
at $\theta$ = 30$^{\circ}$. 
From a detailed analysis by comparison of the spectrum at $\theta$ = 30$^{\circ}$ 
with the results of the band-structure calculation, 
we have estimated the 5d weight to the total density of state 
at $E_F$ as 40-60\%. 
On the other hand, we can estimate the 5d contribution from an analysis of 
$I_{\theta = 90^{\circ}}/I_{\theta = 30^{\circ}}$ in which the relative photoelectron 
cross-sections and angular distributions are taken into account (see Appendix). 
It should be noted that we do not need to use any information obtained 
from the band-structure calculation for this analysis. 
We have successfully estimated the 5d contribution in gold as $50\pm30$\% 
from the analysis of $I_{\theta = 90^{\circ}}/I_{\theta = 30^{\circ}}$ 
while the precise estimation is rather difficult because of possible deviations 
of the actual intensity ratios and cross-sections from the calculated values. 
The mutually consistent results from these two independent analyses 
undoubtedly indicate the prominent 5d contribution to the joint density of 
states in the energy range of $E_F-$2 eV in gold.

\section{Discussions}

The Fermi surface topology of the noble metals, 
which reflects a nature of the conduction electrons, 
is partially deviated from that expected for free electrons 
in crystalline solids~\cite{dHvAAg,dHvAAu}. 
This has been shown theoretically to be resulting from the hybridization 
of the $(n+1)$sp band with the $n$d bands~\cite{Auband1,Agband1,Jepsen1}. 
It has also been predicted that the d-sp hybridization near $E_F$ is 
quantitatively stronger for gold than for silver due to different energy and 
different degree of itinerancy between the 5d and 4d bands. 
Indeed, the experimentally observed threshold of the ``$n$d bands" by 
our and previous~\cite{PES1970,HAXPESAg} experiments is closer to 
$E_F$ for gold ($\sim$2 eV) than for silver ($\sim$4 eV) as shown 
in Figures~\ref{AgVal}~and~\ref{AuVal}. 
The strong itinerancy and d-sp hybridization for the 5d orbitals in gold 
are due to the relativistic effects as discussed for the long 
time~\cite{Auband2,Auband1,Jepsen1,MattheissPt}.
Our findings, the prominent 5d contribution to the conduction electrons in gold 
and essentially negligible 4d mixture for silver, 
are understood as the results of the markedly different strength of 
d-sp mixing near $E_F$. 
On the other hand, it should be noted that 
such a qualitative difference of the d mixture 
in the conduction band crossing $E_F$ has not been predicted 
from the band-structure calculations, 
in which additional electron correlation effects are not taken into account. 

The stability or reactivity of solids in air has directly been discussed 
in terms of kinetic, thermodynamic and surface properties etc. 
by using a Born-Haber cycle, 
which are determined by such fundamental physical properties 
as crystal structure, lattice parameter, the nominal number of valence 
and conduction electrons, electronic dispersion near $E_F$, 
Fermi surface topology and orbital symmetry of conduction electrons. 
On the other hand, the reactivity of the noble metals can also be 
discussed from a viewpoint of the electronic structure as seen 
in ref.~\onlinecite{Hammer}. 
%
Our results experimentally reveal one of few decisive differences 
in such properties as mentioned above 
between gold and silver. Namely, the conduction electrons in silver 
with the predominant 5sp character are more free-electron-like than 
those in gold, in which the 5d character is strongly mixed. 
This different orbital contribution to 
the conduction electrons could be related to the different stability 
in air between gold and silver. 
In addition to our results, it is known that platinum with fcc structure and 
nominal 5d$^9$6sp$^1$ configuration is also very stable in air, 
for which a whole valence-band dispersions are predicted to be almost 
the same as those for gold except for the position of $E_F$ 
and therefore the 5d contribution to the conduction electrons is 
undoubtedly predominant~\cite{MattheissPt}. 
Considering this fact and our results, we can conclude that 
the prominent 5d contribution to the conduction electrons has a role 
to protect gold from the oxidation in air 
in addition to the preceding discussion~\cite{Hammer}, 
where the 5d orbitals are thought to be partially bound to the nucleus sites 
but considerably itinerant due to the relativistic effects. 
If the 5d contribution were negligible, 
gold would tend to be gradually oxidized in air as silver does. 
We note that a similar discussion of reactivity in terms of 
the d mixing in the conduction electrons is not applicable for copper 
even if the 3d contribution would be experimentally verified 
since the character of the 3d orbital is quite different from 
that of the 5d orbital. 
Indeed, both nickel (nominally 3d$^8$4sp$^2$ or 3d$^9$4sp$^1$) and 
zinc (nominally 3d$^{10}$4sp$^2$) are not very stable in air. 

We refer to possible other effects of the prominent 5d contribution 
to the conduction electrons in gold. 
Our result has revealed the presence of the intrinsic 5d holes in gold, 
which have previously been proposed from the results of the band-structure 
calculations~\cite{Jepsen1,MattheissPt} 
and the L$_{2,3}$-edge x-ray absorption~\cite{AuXAS} 
although these studies have not given clear evidence of the 5d holes. 
[It should be noted that the 2p-6s and/or 2p-$k$d ($k > 5$) transitions 
also take place in addition to the 2p-5d transition in the 
L$_{2,3}$-edge absorption.] 
It is expected that there are 5d holes even in recently investigaed 
gold nanoparticles showing magnetism~\cite{MagAu} 
due to the strong d-sp hybridization. 
The 5d holes will increase if the conduction electrons are transferred 
from the nanoparticles to neighbouring molecules as discussed 
in reference~\onlinecite{MagAu}. 
In such a situation, it is natural to consider a role of the 5d holes 
for the magnetism in the gold nanoparticles although our result does not 
give direct evidence of the 5d magnetism. 

\begin{figure}
\begin{center}
\includegraphics[width=8cm,clip]{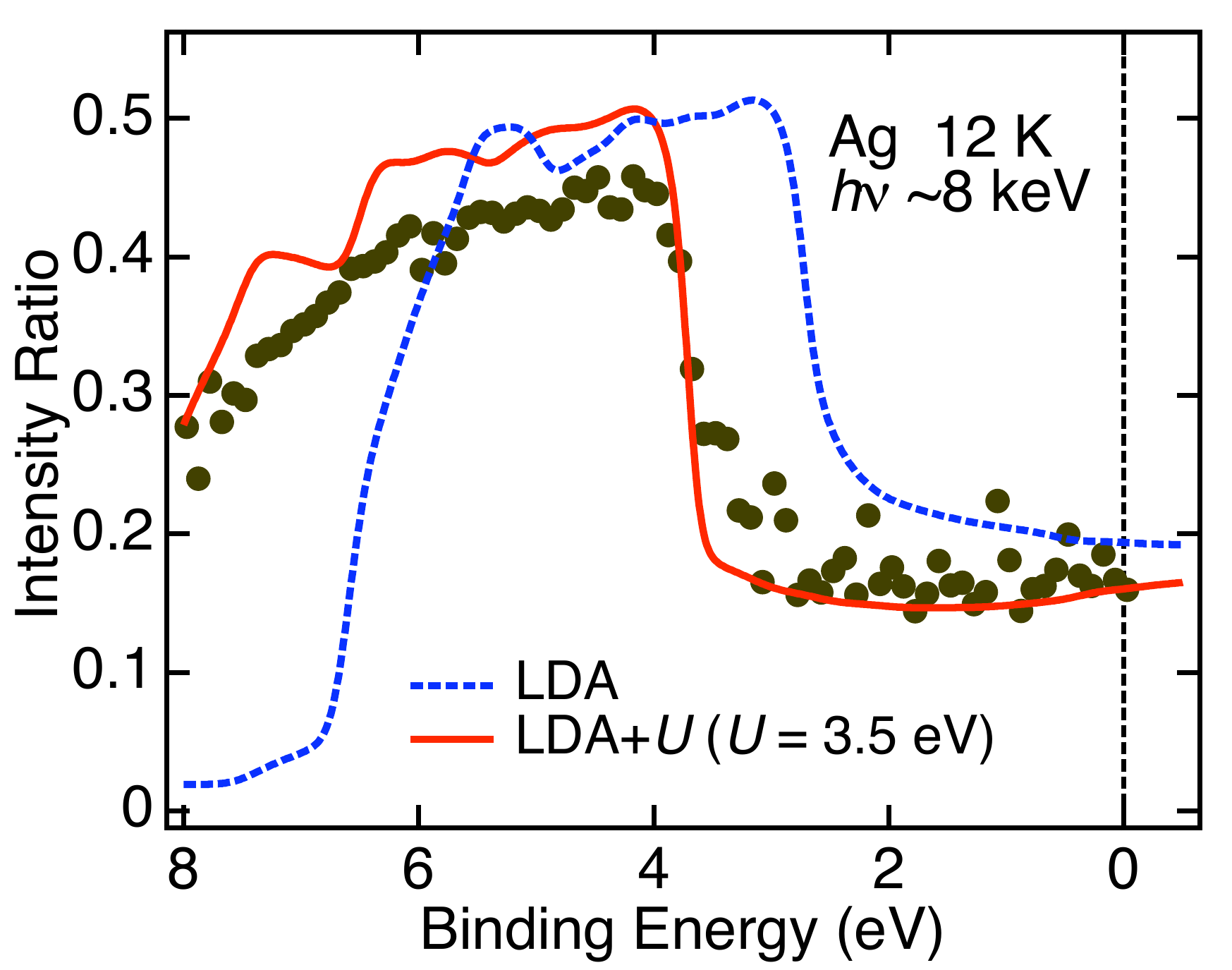}
\end{center}
\caption{
Comparison of the experimental intensity ratio 
$I_{\theta = 90^{\circ}}/I_{\theta = 30^{\circ}}$ (dots) 
for silver with the predicted ratio 
from the band-structure (LDA) calculation 
by using a linear muffin tin orbital method, 
and that from 
the LDA+$U$ calculation in which an on-site Coulomb 
interaction value $U =$ 3.5 eV is chosen. 
For the calculated ratios, 
the cross-sections and angular distributions 
at $h\nu \sim$8 keV~\cite{Pdep1,Pdep3} are taken into account.  
}
\label{AgCompare}
\end{figure}

The band-structure calculations, in which all the valence-band 
electrons are treated as itinerant ones, basically give the results 
of conduction electrons with noticeable d-sp mixing 
for silver and gold. 
Such a calculated result well explain the experimental 
spectra of gold. 
For silver, the experimental 4d contribution in the spectra near $E_F$ 
is much less than that from the calculation. 
This deviation can be understood by considering the 
finite 4d electron correlation effects in silver. 
It is naturally expected that the correlated orbital 
contribution to the conduction electrons is suppressed 
due to the localization 
when the on-site Coulomb interactions are switched on, 
as seen for many rare-earth compounds. 
In order to verify whether the above scenario is correct for silver 
or not, we have also performed the LDA+$U$-like~\cite{LDAU97}  
calculation by using the linear muffin-tin 
orbital method~\cite{LDA1}, in which the 
on-site Coulomb interaction value $U$ = 3.5 eV is 
applied to the 4d orbitals.
Figure~\ref{AgCompare} shows the comparison of 
the ratio $I_{\theta = 90^{\circ}}/I_{\theta = 30^{\circ}}$ 
for silver with the predicted ratios from the calculations. 
One can notice that the LDA+$U$ results 
better explain the experimental ratio near $E_F$  
as well as in a wide valence-band region than 
the LDA calculation. 
We thus conclude that the 4d electron correlation 
effects are responsible for its negligible contribution 
to the conduction electrons in silver. 

\section{Conclusions}

In conclusions, we have performed 
the polarization-dependent hard x-ray photoemission 
for gold and silver, from which the prominent 5d and 
negligible 4d contributions to the conduction electrons 
in gold and silver have been revealed. 
The 4d electron correlation effects are found to be 
essential for the negligible 4d contribution in silver. 

\section*{Acknowledgements}

We thank M. Suzuki, Y. Komura, Y. Tanaka, Y. Nakatsu, G. Kuwahara, A. Yamasaki, 
K. Mima, Y. Miyata, R. Yamaguchi, K. Terashima, T. Yokoi, 
S. Kitayama and K. Kimura for supporting the experiments. 
We are grateful to H. Tada, T. Kimura and T. Nakamura for discussions. 
This work was supported by Grant-in-Aid for Scientific Research (21740229, 21340101), 
Bilateral Programs with Russia, Innovative Areas (20102003) and 
the Global COE (G10) from MEXT and JSPS, Japan. 
IN thanks RFFI grants 08-02-00021, 08-02-91200, 
and grant of President of Russia MK-614.2009.2.

\appendix
\section{Estimation of the 5d contribution to the conduction electrons in gold from the spectra}

The intensity ratio $I_{\theta = 90^{\circ}}/I_{\theta = 30^{\circ}}$ in the vicinity 
of $E_F$ shown in the upper panel of Figure~\ref{AuVal} (a), defined as $y$, 
is experimentally obtained as $y = 0.22\pm0.02$. 
The ratio for the 5d$_{5/2}$ states is calculated as 0.3 by using the parameters 
in references~\onlinecite{Pdep2,Pdep3}. 
We have calculated the ratio at the photoelectron kinetic energies 
of 1, 3, 5 and 10 keV and then obtained the above value at 8 keV 
by the interpolation 
since there are no calculation parameters for the 8-keV photoelectrons. 
As shown in Figure~\ref{AuVal} (a), the calculated value for the 5d$_{5/2}$ 
excitation is very close to the experimentally estimated ratio 
at the binding energy of 2-5 eV, 
which corresponds to the 5d-band region. 
$I_{\theta = 90^{\circ}}/I_{\theta = 30^{\circ}}$ for the 6sp band is 
assumed to be 0.1 
because this value is very close to the value calculated 
(by the same manner as that for the 5d state described above) 
for the (Tl) 6p state and that for the 5s core-level excitation 
in our experiment (Figure~\ref{AuCore}). 
An photoemission intensity ratio per an electron of the 5d to 6sp states 
at $\theta$ = 30$^{\circ}$ and $h\nu$ = 8 keV defined as $a$, 
which has been expressed as a ratio of products of the calculated 
angular distribution at $\theta$ = 30$^{\circ}$ and the photoionization cross-section, 
has been estimated as $a = 1.3-2$. 
This value is not deviated very much from 1, 
therefore the experimental spectra near $E_F$ at $\theta$ = 30$^{\circ}$ 
can be well reproduced by the naive sum of the partial density of states 
as shown in Figure~\ref{AuVal} (b). 
When $x$ is defined as the relative 5d contribution 
to the conduction electrons, and the relative 6sp contribution 
in the vicinity of $E_F$ is set to $1-x$, 
the observed photoelectron intensity in the spectrum 
at $\theta$ = 30$^{\circ}$ is expressed as $ax+(1-x)$. 
Since the spectral weight is reduced at $\theta$ = 90$^{\circ}$ 
as the factor of 0.3 (0.1) for the 5d (6sp) states, 
the ratio $y$ is expressed as $y = \{0.3ax+0.1(1-x) \}/\{ax+(1-x)\}$. 
Then we can estimate the 5d contribution $x$ 
as $\sim$0.5 when $a = 1.5$ and $y = 0.22$.

\end{document}